\documentstyle[preprint,aps]{revtex}
\begin{document}
\begin{center}
{\bf R. Beck and H.-P. Krahn: Constraining the $(\gamma, \pi)$
amplitude for $E2$ }$N \rightarrow \Delta$
\end{center}

In a recent Letter \cite{beck}
we have reported precision measurements
of differential cross sections and polarized photon asymmetries for the
reaction $\vec{\gamma}p$ $\rightarrow$ $p\pi^0$ with the DAPHNE--detector, 
using tagged photons at the Mainz Microtron MAMI.
The above Comment\cite{andy} critizies our value $R_{EM} =
ImE^{3/2}_{1^+} / ImM^{3/2}_{1^+} = -(2.5 \pm 0.2 \pm 0.2)\% $, because of
possible ambiguities stemming from contributions of higher partial waves.

\begin{figure}
\centerline{\psfig{figure=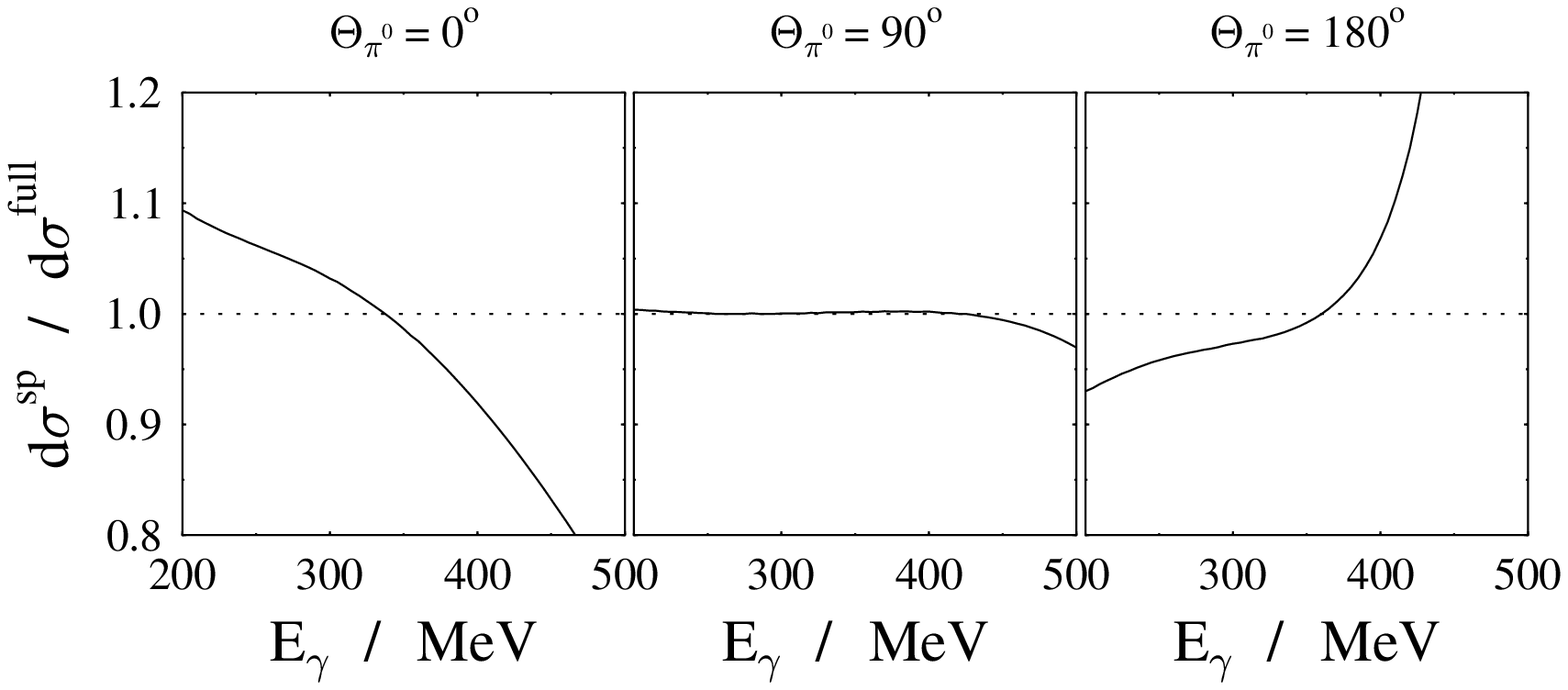,width=16.0cm,height=8.0cm}}
\caption{
The ratio of the differential cross section for 
only $s$-- and $p$--waves contributions to the cross section where 
higher partial waves have been taken into account 
(truncation at $f$--waves, Born 
contribution for $l_{\pi} \ge 4$)  VPI[SM95].}
\label{fig:saidsp}
\end{figure}

We are using Eqs.~(3) to (7) in our paper \cite{beck} to extract the $R_{EM}$
value. These Eqs. are exact under the assumption that only s-- and p--waves
contribute. To study the validity of this assumption we investigated the
effects of higher partial waves ($l_{\pi} \ge 2$) in Eqs.~3 to 7. 
The inclusion of d--waves results in a modification of Eq.~(3) to
\begin{eqnarray}
\frac{d\sigma}{d\Omega} = \frac{q}{k}( A + B cos(\theta) + C cos^2(\theta) +
D cos^3(\theta) + E cos^4(\theta)) ~~. \; 
\end{eqnarray}
Two additional coefficients D and E appear and furthermore the
coefficients A, B and C are modified according to
\begin{eqnarray}
A &\simeq& A(s_{wave},p_{wave}) +  {\rm Re} \left[E_{0+}d^*\!\!_{wave}  \right] +  
| d_{wave} |^2  \; , \\
B &\simeq& B(s_{wave},p_{wave}) + {\rm Re} \left[ (M_{1+}-M_{1-}) d^*_{wave}  \right] \; , \\
C &\simeq& C(s_{wave},p_{wave}) + {\rm Re}\left[ E_{0+} d^*\!\!_{wave} \right] + |
d_{wave} |^2 \; , \\
D &\simeq& {\rm Re} \left[ (M_{1+}-M_{1-}) d^*\!\!_{wave} \right] \; , \\
E &=& | d_{wave} |^2 \; , 
\end{eqnarray}
where $s_{wave},~p_{wave}$ and $d_{wave}$ are combinations of the
corresponding partial wave multipoles.
The effect is largest for the coefficients B and D, where an interference term
between the large $M_{1+}$ and the d--waves occurs. But at the top of the
resonance ($\delta_{33} = 90^0$) the contributions of these terms can be 
neglected, e.g.
\begin{eqnarray}
{\rm Re} \left[ (M_{1+}-M_{1-}) E_{2-}^* \right] &=& 
{\rm Re} (M_{1+}-M_{1-}) {\rm Re} E_{2-} + 
{\rm Im} (M_{1+}-M_{1-}) {\rm Im} E_{2-} ~~~.  \;
\end{eqnarray}
The first term vanishes, because ${\rm Re} (M_{1+}-M_{1-})$ goes through zero
near the resonance energy ($E_{\gamma} = 340$ MeV) and the second term can be
neglected, because ${\rm Im} E_{2-}$ is small due to a phase close to zero.
Fig.~\ref{fig:saidsp} shows the ratio of the differential cross section for 
only $s$-- and $p$--waves contribution to the cross section where higher 
partial waves have been taken into account (truncation at $f$--waves, Born 
contribution for $l_{\pi} \ge 4$)  VPI[SM95]\cite{Wor96}. The ratio is shown 
at $\theta_{\pi} = 0^0,~90^0$ and $180^0$ in the energy region of $200$ to $500MeV$. 
At $\theta_{\pi} = 90^0$, the contributions from the higher partial waves are
far below $1\%$, since there is only an interference term with the $s$--wave
$E_{0^+}$  (e.g. $Re(E_{0^+}d^*_{wave})$).
Below and above the resonance, however, contributions from $l_{\pi} \ge 2$ are 
of the order of $10 - 20 \%$ of the differential cross section at 
$0^0$ and $180^0$.
This will affect the $C_{\|}$--coefficient below and above the resonance.

\begin{figure}
\centerline{\psfig{figure=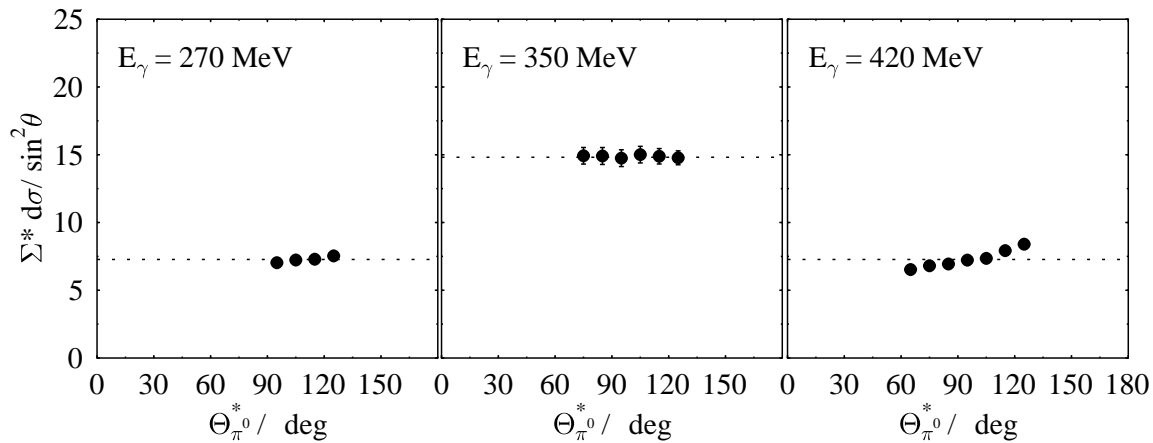,width=16.0cm,height=7.5cm}}
\caption{The linear polarization cross section difference
$d\sigma_{\bot} - d\sigma_{||} = \Sigma d\sigma / sin^2\theta$
for $p(\vec{\gamma},p)\pi^0$.}
\label{fig:sp}
\end{figure}

Another observable which is sensitive to a contribution of higher partial
waves is the linear polarization cross section difference 
$d\sigma_{\bot} - d\sigma_{||}$.
Fig.~\ref{fig:sp} shows $d\sigma_{\bot} - d\sigma_{||}$ in a power series 
expansion in $cos\theta$ for our $(p,\pi^0)$ data
\begin{equation}
d\sigma_{\bot} - d\sigma_{||} = \Sigma d\sigma / sin^2\theta = \frac{q}{k}(A_{\Sigma} +
B_{\Sigma} cos\theta + C_{\Sigma} cos^2\theta)
\end{equation}
with
\begin{eqnarray}
A_{\Sigma} &\simeq& A(s_{wave},p_{wave}) +  {\rm Re} \left[E_{0+}d^*\!\!_{wave}  \right] +  
| d_{wave} |^2  \; , \\
B_{\Sigma} &\simeq& {\rm Re} \left[ (M_{1+}-M_{1-}) d^*_{wave}  \right] \; , \\
C_{\Sigma} &=& | d_{wave} |^2 \; . 
\end{eqnarray}
In the case, where only $s$-- and $p$--waves contribute, this 
difference should be equal to $A_{\Sigma}$ and therefore 
constant, independent of the pion angle $\theta_{\pi}$. 
The $B_{\Sigma}$--coefficient is an
interference term between the large $M_{1^+}$--amplitude and the $d$--waves. 
There are {\it NO} ambiguities and {\it NO}  indications
for a non--Born contribution for higher partial waves $l_{\pi} \ge 2$ around 
the $\Delta(1232)$--resonance in our $(p,\pi^0)$ data.  

In Table 1 of reference \cite{andy}, we believe that the LEGS analysis is
running into the classical problem of a multipole analysis: How to handle
systematic errors coming from different experiments? It is certainly not 
reasonable to increase the number of partial waves until the fit is stable,
because there is already systematics absorbed into the partial waves.
One has instead to look at observables, which are sensitive to the d--wave
contribution.  This has already been pointed out by the original multipole
analysis of the Khark'hov data\cite{Gru81}. In this work it was
demonstrated, that there is no need of higher partial waves (non--Born
contribution) in the polarization observables ($\Sigma$, $T$ and $P$).
There is, as correctly pointed out in\cite{Gru81}, a definite incompatibility
of the experimental Bonn data on $d\sigma$ at the extreme forward and 
backward angles to the photon asymmetry result $\Sigma(90^0)$ from Khark'hov. 
This is very important, because these two observables have a similar 
$M_{1^+}E_{1^+}$ interference term in this angular range.

In conclusion, there are {\it NO} ambiguities stemming from necglecting
contributions of higher partial waves in our $(p,\pi^0)$ analysis and there
is {\it NO} reason to change our value $R_{EM} =
ImE^{3/2}_{1^+} / ImM^{3/2}_{1^+} = -(2.5 \pm 0.2 \pm 0.2)\% $.

R. Beck and H.-P. Krahn \\
{\it
 Institut f\"ur Kernphysik der Universit\"at Mainz,\\
 Becherweg 45, 55099 Mainz, Germany}

\input{psfig}
\end{document}